# On free-stream preservation in stationary grids for arbitrary linear upwind schemes

**Qin Li, Dong Sun, Hanxin Zhang**

State Key Laboratory of Aerodynamics, Mianyang, Sichuan, 621000, China,

and

National Laboratory of Computational Fluid Dynamics, Beijing University of Aeronautics and Astronautics, Beijing 100191, China

**Abstract** In order to improve the application maturity of high-order difference schemes, the free-stream preservation property, whose importance has been widely recognized in recent years, has been developed into a focus of study.. In past literatures, only central schemes are considered to be suitable for free-stream preservation. In this study, the methodology for arbitrary linear schemes to achieve the property is investigated. First, derivations of grid metric by Thomas, Lombard and Neier (AIAA J., 17, 10, 1978 and J. Spacecraft and rocket, 27, 2, 1990) are reviewed, through which linear schemes for the metric and unsplit flux could attain the property by the proof of Vinokur and Yee firstly (NASA TM 209598, 2000). In practical applications, flux splittings are usually at presence and therefore the direct use of upwind schemes seems difficult to fulfill free-stream preservation. To overcome the difficulty, two attempts are made: firstly, a central-scheme-decomposition is worked out, through which a central difference scheme is derived to approximate the first-order partial derivative in metric evaluations; secondly, treatments are proposed for flux splitting and a concrete example is presented. Through these two attempts, the linear upwind node schemes can achieve the property. For half-node or mixed type schemes, interpolations should be used to derive variables at half nodes. As a result, a directionally consistent interpolation is proposed, which is shown to be necessary in order to avoid violation of the metric identity and free-stream preservation. Two numerical problems are also tested, i.e., the free-stream and vortex preservation on wavy, largely randomized and triangular grids. Numerical results validate aforementioned theoretical outcomes; especially, simulations on the triangular grids indicate the methods discussed in this study, which are typical algorithms on structured grids, has the application potential for problems on unstructured meshes.

## 1. Introduction

It is well-known in computational fluid dynamics (CFD) practices that the use of deformed grids leads to unsatisfactory results. When this situation occurs, one usually prefers to improve the grid quality rather than inquire into the reason, and such efforts always work well when using TVD schemes and finite-volume methods. However, progress seems to be slow on utilizing high-order difference schemes to solve complex problems. In Ref. [1], Visbal and Gaitonde demonstrated distinct errors caused by metric evaluations when using high-order schemes, and "these errors can catastrophically destroy the fidelity of the approaches". Later, Nonomura, Lizuka and Fujii [2] numerically investigated similar problems by using two concrete schemes on deformed grids, which verified again the importance of the metric computation. Through their work, a topic with long history regarding metric-generated error was re-brought into the sight of CFD community.

At least in 1974, Vinokur [3] gave the conservative forms of Euler Equations in stationary curvilinear coordinate systems, which was accomplished in tensor description after coordinate transformation. Although the tensor form might be unfamiliar to current CFD practitioners, the acquisition of the conservative equations indicates the use of theoretically zero-valued terms, i.e., metric identities. In 1978, Pulliam and Steger [4] pointed out that the presumed zero-valued identities might actually have non-zero value in computations. Hence when the uniform flow condition is imposed, the flow field might change and so-called free-stream preservation (*FSP*) property could be broken. To overcome the difficulty, they proposed an averaging procedure for remedy in the second-order difference frame-work. In the meanwhile, Thomas and Lombard [5, 6] discussed problems with moving grids and proposed another conservative relation regarding the change of grid metrics and transformation Jacobian, which was called as "geometric conservation law" (GCL). They showed that GCL might also be violated if special treatment was absent. Afterwards, series of studies have been made on errors generated by metric/Jacobian evaluations, and theoretical outcomes were obtained thereafter. In this study, investigations concentrate only on metric evaluations in the case of stationary grids.

Next, a brief discussion about the history and achievements of eliminating errors generated in metric evaluations is summarized into the following three aspects:

(1) The evolution of the metric form. In CFD literatures, grid metrics are usually shown in products of individual derivatives of coordinates, e.g., $\hat{\xi}_x = y_\eta z_\zeta - z_\eta y_\zeta$. When this form is chosen, it seems that only second-order schemes with averaging technique [4] can achieve metric identity (*MI*) or metric cancellation. By using the production rule of derivatives, Thomas and Lombard [6] proposed another "conservative" form of the metric evaluation. By using the form, the restriction of using specific difference scheme to achieve metric cancellation was largely released. In 1990, Thomas and Neier [7] recast the conservative form into a more symmetric one, which was later referred by Vinokur and Yee [8] as "coordinate invariant form".

(2) The idea of using the same scheme for the metric and flux approximation in fluid governing equations. Despite of its unpopularity, this idea actually has existed for a long time. In Ref. [9], Thompson et al. mentioned "numerical evaluation of the metric coefficients by the same difference representation used for the function whose derivative is being represented is preferable over exact analytical evaluation". Gaitonde and Visbal [10] further stated that metrics "computed with the same scheme as employed for the fluxes" can "reduces the error on stretched meshes". In Ref. [10, 11], they showed the use of the sixth-order compact scheme can engender better results than lower order schemes on various deformed grids.

(3) Approaches to avoid errors by grid metrics. After numerically testing various center schemes with an order from the second to the sixth, Gaitonde and Visbal [10] found the coupling of the same-scheme idea with the conservative form of metrics in Ref. [6] can reduce metric-related errors to machine zero. Having observed this result, Vinokur and Yee [8] made analysis and indicated the key lay in the numerical commutativity of the mixed partial derivative. By using the notion of tensor product, they showed an analytically proof on the commutativity if either of two conservative forms [6, 7] and the same discretizing scheme was used. Thus far, all theoretical problems have been settled. Ten years later, investigations were made on the same topics again after noticing pervious works [12-14]. Although analysis with different considerations are conducted [12, 14-16], it is definite that Thomas, Lombard and Neier [6, 7] are credited for the proposition of possible conservative forms of metrics, and Vinokur and Yee [8] firstly gave the proof that the use of same schemes combined with the conservative form can guarantee commutativity of mixed partial derivatives and *MI* accordingly. Besides the above methods by automatic cancellations of metric errors, other effort was observed by positively removing the errors introduced during the equation transformations. At least in Ref. [9], Thompson et al. proposed "the effects of the spurious source terms can be partially corrected, ... by subtracting the product of the metric identities with either a uniform solution or the local solution". The same idea was mentioned by Vinokur [17]. Cai and Ladeinde [18] showed a numerical practice of this regard. It is usually understood that such efforts can alleviate the error to some extent but cannot fully solve the problem.

One of the outcomes of above investigations is that linear central schemes could make *MI* valid and *FSP* achievable when combined with the conservative metrics, e.g., the node or half-node type compact schemes by Lele [19]. As shown in Ref. [1], due to zero dissipation, central schemes cannot work independently in practical problems unless the filters are combined with the schemes. Naturally, it is of interest to know if upwind schemes could achieve *FSP* in stationary grids as well. Thompson et al. [9] indicated complexities would arise by the "flux-vector-splitting", and such procedures will definitely be used by upwind schemes. Although Ref. [12] proposed a delta form of flux difference for this regard, the attempt should not be considered as a general solution. It was widely doubted that whether upwind schemes could achieve *FSP* [2, 12].

In this study, a decomposition approach is proposed for arbitrary upwind schemes, through which the requirement for flux splitting is derived and *FSP* in stationary grids can be realized correspondingly. In Section 2, metric evaluation methods to achieve metric cancellation and *FSP* are first reviewed. In Section 3, analysis is made on metric cancellation and *FSP* regarding linear upwind schemes and corresponding methods are proposed. In Section 4, numerical validations are provided to show the validity of the proposed methods. At last, conclusions are drawn in Section 5.

## 2. Metric identities and evaluations of grid metrics and Jacobian

### 2.1. Metric identities and free-stream preservation

The non-dimensional Navier-Stokes equation in the Cartesian coordinate system is considered as follows:

$$\partial_t Q+\partial_x\left(E-E_v\right)+\partial_y\left(F-F_v\right)+\partial_z\left(G-G_v\right)=0\,. \tag{1}$$

In this equation, $Q=(\rho, \rho u, \rho v, \rho w, e)$, $e=p/(\gamma-1)+1/2\times\rho(u^2+v^2+w^2)$; $E$, $F$ and $G$ are inviscid fluxes and $E_v$, $F_v$, and $G_v$ are viscous ones. The definitions of the fluxes are easy to find in CFD books and will not be repeated here. For a uniform flow, all spatial derivatives in Eq. (1) will be zero and $Q$ will not change. Hence, the property of *FSP* is naturally established.

To solve Eq. (1) in the curvilinear coordinates system, the transformation from the Cartesian coordinates is employed: $(x, y, z)\rightarrow(\xi, \eta, \gamma)$. For the sake of simplicity, some convention of the tensor analysis will be used as the following: $\xi^j$ is for $(\xi, \eta, \gamma)$, $x_i$ is for $x$, $y$ and $z$, and $u^i$ is for $u$,

$v$, and $w$. Using the chain law $\partial_{x_i} = \xi^j_{x_i} \cdot \partial_{\xi^j}$, Eq. (1) becomes

$$\partial_t \hat{Q} + \partial_\xi \left( \hat{E} - \hat{E}_v \right) + \partial_\eta \left( \hat{F} - \hat{F}_v \right) + \partial_\zeta \left( \hat{G} - \hat{G}_v \right) = -(\hat{E} - \hat{E}_v, \hat{F} - \hat{F}_v, \hat{G} - \hat{G}_v) \cdot \vec{I} \,. \tag{2}$$

In the equation, $\hat{Q} = J^{-1} Q$, $J^{-1} = \left| \frac{\partial(x,y,z)}{\partial(\xi,\eta,\zeta)} \right|$, $\hat{E} = (\hat{\xi}_x E + \hat{\xi}_y F + \hat{\xi}_z G)$ with $\hat{\xi}^i_{x_i} = J^{-1} \xi^i_{x_i}$ and $\hat{F}$, $\hat{G}$, $\hat{E}_v$, $\hat{F}_v$, $\hat{G}_v$ can be derived similarly; $\vec{I}$ stands for a vector $\left( I_x, I_y, I_z \right)$ with the member as

$$I_{x_i} = \left( \hat{\xi}_{x_i} \right)_\xi + \left( \hat{\eta}_{x_i} \right)_\eta + \left( \hat{\zeta}_{x_i} \right)_\zeta \equiv \left( \hat{\xi}^j_{x_i} \right)_{\xi^j} . \tag{3}$$

Using the notation $\left( \cdot \right)_{\vec{r}} = \left( \partial_x, \partial_y, \partial_z \right)$, $\hat{\xi}^i_{\vec{r}}$ can be derived as

$$\hat{\xi}^i_{\vec{r}} = \vec{r}_{\xi^j} \times \vec{r}_{\xi^k} \quad , \tag{4}$$

where the indices ($i$, $j$, $k$) are cyclic. It should be noticed the outside derivative $\partial_{\xi^j}$ in Eq. (3) is from the fluid dynamic operations in Eq. (2), while similar ones inside $\hat{\xi}^i_{\vec{r}}$ originate from the metric computation. It is trivial due to the commutativity of partial differential derivatives, $\vec{I} = 0$ or the metric identity holds, and the popular conservative form will be established by discarding the right-hand side term of Eq. (2). If numerical schemes are used to evaluate the spatial derivatives, the equation can be further written as

$$\partial_t \hat{Q} + \delta_\xi \left( \hat{E} - \hat{E}_v \right) + \delta_\eta \left( \hat{F} - \hat{F}_v \right) + \delta_\zeta \left( \hat{G} - \hat{G}_v \right) = 0 \,, \tag{5}$$

where $\delta$ denotes the numerical approximation of $\partial$, e.g., various difference schemes. Eq. (5) is the most common choice by simulations, while Eq. (2) is also observed as being the governing equation for discretization [18]. No matter what kind of equations are used, no change should be aroused for flow variables when the uniform-flow condition is imposed. Under this condition, the following equation is referred in literatures [12]:

$$\partial_t \hat{Q} + \left( E_\infty, F_\infty, G_\infty \right) \cdot \vec{I}^* - \left( E_{v,\infty}, F_{v,\infty}, G_{v,\infty} \right) \cdot \vec{I}^* = 0 \,, \tag{6}$$

where $\vec{I}^*$ is the numerical evaluation of $\vec{I}$ by replacing $\partial$ by $\delta$ correspondingly, and the subscript "$\infty$" denotes the uniform-flow state. From Eq. (6), the establishment of *FSP* seems equal to the numerical validity of *MI*, which is consistent to the discard of $\vec{I}$ in Eq. (2). However, when the flux splitting is considered, Eq. (6) usually cannot be attained for all numerical schemes, therefore the separate validity on *MI* does not promise *FSP* sufficiently. The acquisition of validity of Eq. (6) will be discussed in Section 3.

## 2.2. The evaluation of metrics to achieve metric identities

It has been indicated by Ref. [5, 10] that the original form of $\hat{\xi}^i_{\vec{r}}$ by Eq. (4) is hard to achieve

*MI* by using general schemes, especially the high-order ones. Using the production rule of derivatives, Thomas and Lombard [5] first proposed the equivalent conservative form as:

$$\hat{\xi}^i_{x_{i'}} = \left[\left(x_{j'}\right)_{\xi^j} \cdot x_{k'}\right]_{\xi^k} - \left[\left(x_{j'}\right)_{\xi^k} \cdot x_{k'}\right]_{\xi^j} \quad , \tag{7}$$

where two sets of indices with and without prime are cyclic. Replacing Eq. (4) with Eq. (7) , they stated that $\vec{I}$ can "vanish identically when central difference operators are used to evaluate the spatial derivatives", and the limitation about specific technique by Pulliam and Steger [4] was relieved. Besides, it can be seen that the positions of $x_{j'}$ and $x_{k'}$ in Eq. (7) are not equal. Possibly noticing this unbalance, Thomas and Neier [7] further proposed the following symmetric form:

$$\hat{\xi}^i_{\vec{r}} = \frac{1}{2}\left[\left(\vec{r} \times \vec{r}_{\xi^k}\right)_{\xi^j} + \left(\vec{r} \times \vec{r}_{\xi^j}\right)_{\xi^k}\right], \tag{8}$$

which is actually the average of Eq. (7) and its reciprocal form: $\left[\left(x_{k'}\right)_{\xi^k} \cdot x_{j'}\right]_{\xi^j} - \left[\left(x_{k'}\right)_{\xi^j} \cdot x_{j'}\right]_{\xi^k}$. Later on, Eq. (8) was referred to by Vinokur and Yee [8] with stress on its coordinate invariant property.

In the following, Eq. (7) is used to illustrate why Eq. (7) and (8) might yield metric cancellation, while the mechanism is the same for Eq. (8). It is obvious that from Eq. (7), the second term of $\hat{\xi}^k_{x_{i'}}$ is $-\left[\left(x_{j'}\right)_{\xi^j} \cdot x_{k'}\right]_{\xi^i}$, therefore its partial derivative with $\xi^k$ will cancel out the partial derivative of the first term in $\hat{\xi}^i_{x_{i'}}$ with $\xi^i$ when considering $\partial_{\xi^j \xi^k} = \partial_{\xi^k \xi^j}$. By using similar operations, the metric cancellation can be attained. If the same property could also be possessed by difference schemes, *MI* should be numerically achieved. Apparently Vinokur and Yee [8] noticed this critical point and then gave a proof, which will be briefly reviewed as follows.

Before further discussion, the tensor or Kronecker product of two arbitrary matrices *A* and *B* is introduced first, which generates a block matrix with the element: $\left(A \otimes B\right)_{ij} = A_{ij}B$. Based on the concept, the mixed product rule exists [8] for two pairs of conformable matrices {A, C} and {B, D} as:

$$\left(A \otimes B\right)\left(C \otimes D\right) = AC \otimes BD, \tag{9}$$

where $AC$ denotes ordinary matrix product. Then consider the 3-D curvilinear coordinates system with the dimension ($l$, $m$, $n$) in ($\xi$, $\eta$, $\zeta$) directions. Take $\xi$ direction as an example. Suppose the difference scheme for $u_\xi$ can be generally expressed as $A^\xi u_\xi = B^\xi u$, where $A^\xi$ and $B^\xi$ are $l$ by $l$ matrices, and $u_\xi$ and $u$ are $l$-dimensional vectors. Assuming the computational order for the whole discrete variables is in the sequence of $\xi$, $\eta$ and $\zeta$, then the equation for all $u_\xi$ can be written as $\bar{A}^\xi \bar{u}_\xi = \bar{B}^\xi \bar{u}$, where $\bar{u}$ and $\bar{u}_\xi$ are $l \times m \times n$-dimensional vectors of $u$ and $u_\xi$, and $\bar{A}^\xi$ and $\bar{B}^\xi$ are ($l \times m \times n$) by ($l \times m \times n$) matrices with the form

$$\begin{cases} \overline{A}^{\xi} = I^{n} \otimes \left( I^{m} \otimes A^{\xi} \right) \\ \overline{B}^{\xi} = I^{n} \otimes \left( I^{m} \otimes B^{\xi} \right) \end{cases}. \tag{10}$$

In Eq. (10), $I^n$ is $n$ by $n$ identity matrix and similarly does $I^m$. In the same way, the equation for all $u_\eta$ can be derived as $\overline{A}^{\eta}\overline{u}_{\eta} = \overline{B}^{\eta}\overline{u}$ with

$$\begin{cases} \overline{A}^{\eta} = I^{n} \otimes \left( A^{\eta} \otimes I^{l} \right) \\ \overline{B}^{\eta} = I^{n} \otimes \left( B^{\eta} \otimes I^{l} \right) \end{cases}. \tag{11}$$

Then the discretization of the mixed derivative $u_{\xi\eta}$ becomes: $\overline{A}^{\eta}\left(\overline{A}^{\xi}\overline{u}_{\xi}\right)_{\eta} = \overline{A}^{\eta}\overline{A}^{\xi}\overline{u}_{\xi\eta}$ $= \overline{B}^{\eta}\overline{B}^{\xi}\overline{u}$ . Using Eq. (9), $\overline{A}^{\eta}\overline{A}^{\xi}$ = $\left[I^{n} \otimes \left(A^{\eta} \otimes I^{l}\right)\right]\left[I^{n} \otimes \left(I^{m} \otimes A^{\xi}\right)\right]$ = $I^{n} \otimes \left[\left(A^{\eta} \otimes I^{l}\right)\left(I^{m} \otimes A^{\xi}\right)\right] = I^{n} \otimes A^{\eta} \otimes A^{\xi}$. In the same manner, $\overline{A}^{\xi}\overline{A}^{\eta} = I^{n} \otimes A^{\eta} \otimes A^{\xi}$, and therefore $\overline{A}^{\eta}\overline{A}^{\xi} = \overline{A}^{\xi}\overline{A}^{\eta}$. Similarly, $\overline{B}^{\eta}\overline{B}^{\xi} = \overline{B}^{\xi}\overline{B}^{\eta}$. Hence, $\overline{u}_{\xi\eta} = \overline{u}_{\eta\xi}$ or the numerical commutativity is satisfied. In a similar way, $\overline{u}_{\xi\zeta} = \overline{u}_{\zeta\xi}$ and $\overline{u}_{\eta\zeta} = \overline{u}_{\zeta\eta}$ can be established. More details are suggested to Ref. [8].

As mentioned in Ref. [8], the above proof stands for compact or non-compact schemes, and arbitrary boundary conditions can be incorporated as well. Another implication in the proof is the difference scheme as $A^{\xi}$ is constant and consistent in the evaluation of $u_{\xi\eta}$ and $u_{\eta\xi}$. Although trivial, the implication can be interpreted in the context of CFD as: to achieve commutativity, the scheme in each coordinate direction should be linear and keep the same form for the metrics and flux approximation.

It is conceivable that when the type of schemes degenerates to finite difference, simpler proofs might be available. Such practices can be observed in Ref. [12, 20], and the process is reiterated as follows. Without losing generality, suppose the difference operators $\delta_\xi$ and $\delta_\eta$ at ($i$, $j$, $k$) can be expressed as:

$$\begin{cases} \delta_{\xi}\left(\cdot\right)_{i,j,k} = \dfrac{1}{\Delta}\sum\limits_{i_1=-m_1}^{n_1} a_{i_1}^{\xi}\left(\cdot\right)_{i+i_1,j,k} \\ \delta_{\eta}\left(\cdot\right)_{i,j,k} = \dfrac{1}{\Delta}\sum\limits_{j_1=-m_2}^{n_2} a_{j_1}^{\eta}\left(\cdot\right)_{i,j+j_1,k} \end{cases}, \tag{12}$$

where $\Delta$ denotes spatial interval, then

$$\delta_{\xi\eta}\left(u\right)_{i,j,k} = \frac{1}{\Delta^2}\sum_{j_1=-m_2}^{n_2} a_{j_1}^{\eta}\left(\sum_{i_1=-m_1}^{n_1} a_{i_1}^{\xi}\left(u\right)_{i+i_1,j+j_1,k}\right) = \frac{1}{\Delta^2}\sum_{i_1=-m_1}^{n_1} a_{i_1}^{\xi}\left(\sum_{j_1=-m_2}^{n_2} a_{j_1}^{\eta}\left(u\right)_{i+i_1,j+j_1,k}\right) = \delta_{\eta\xi}\left(u\right)_{i,j,k}$$

In the formula, the trivial commutativity of summation in linear algebra is used.

.

It is worthy of notice that only the consistent use of the linear scheme is required in above proofs, while upwind ones are allowable and schemes can be different in different coordinate direction. In practical applications, the flux will be split into two parts and different upwind schemes will be used correspondingly. In this sense, the fulfill of *MI* does not directly equal to *FSP*.

2.3. The evaluation of Jacobian

Although not related with *MI*, it was reported in literatures [21, 14] that different forms of $J^{-1}$ might influence the level of grid-generated errors on seriously deforming grids. It is the well-known definition for $J^{-1}$:

$$J^{-1} \equiv \left| \frac{\partial(x,y,z)}{\partial(\xi,\eta,\zeta)} \right| = \vec{r}_{\xi^i} \cdot \left( \vec{r}_{\xi^j} \times \vec{r}_{\xi^k} \right), \tag{13}$$

where the indices are cyclic. If terms like $\vec{r}_{\xi^i}$ are computed individually, the circulation of indices in Eq. (13) will not result in different values of $J^{-1}$. Considering Eq. (4), $J^{-1}$ can apparently be expressed as $\vec{r}_\zeta \cdot \hat{\zeta}_{\vec{r}}$, which might be numerically different to $\vec{r}_\xi \cdot \hat{\xi}_{\vec{r}}$ or $\vec{r}_\eta \cdot \hat{\eta}_{\vec{r}}$. So if $\hat{\xi}^{i'}_{\vec{r}}$ is derived by Eq. (7) or (8), the different choice of curvilinear coordinates in Jacobian derivation might yield different $J^{-1}$. Practically speaking, it is reasonable that no warranty for specific choice of coordinate exists which have the least numerical errors, hence it is natural to use average of three candidates as what is employed in Eq. (8). This technique is the one proposed in Ref. [21] and later referred by Ref. [14]. Furthermore, Abe et al. [21] integrated the metric identities into $\vec{r}_{\xi^i} \cdot \hat{\xi}^i_{\vec{r}}$ and obtained the conservative form of $J^{-1}$ as: $\frac{1}{3}\left(\vec{r} \cdot \hat{\xi}^i_{\vec{r}}\right)_{\xi^i}$.

Although the above disposes seem to appear recently, they are actually contained in the simple deduction of some basic formulae already-known before. In Ref. [9], two forms of the divergence of the vector $\vec{A}$ in general coordinate system was shown as,

$$\begin{cases} \nabla \cdot \vec{A} = \left( \sqrt{g}\vec{a}^i \cdot \vec{A}_{\xi^i} \right) \Big/ \sqrt{g} & \text{(14.a)} \\ \nabla \cdot \vec{A} = \left( \sqrt{g}\vec{a}^i \cdot \vec{A} \right)_{\xi^i} \Big/ \sqrt{g} & \text{(14.b)} \end{cases},$$

where $\vec{a}^i$ is the contravariant base vector defined as $\sqrt{g}\vec{a}^i = \vec{r}_{\xi^j} \times \vec{r}_{\xi^k}$ with $\sqrt{g} = \vec{r}_{\xi^i} \cdot \left( \vec{r}_{\xi^j} \times \vec{r}_{\xi^k} \right)$. Eq. (14.a) is the conservative variant of Eq. (14a) by using the identity $\vec{a}^i_{\xi^i} = 0$. Thompson et al. [9] also suggested " the product $\sqrt{g}\vec{a}^i$ may be stored at each point" for usage, therefore Eq. (14) can be re-interpreted by CFD as: $\nabla \cdot \vec{A} = \frac{1}{J^{-1}}\left( \hat{\xi}^i_{\vec{r}} \cdot \vec{A}_{\xi^i} \right)$ or $\nabla \cdot \vec{A} = \frac{1}{J^{-1}}\left( \hat{\xi}^i_{\vec{r}} \cdot \vec{A} \right)_{\xi^i}$. Taking $\vec{A}$ as $\vec{r}$, the following result is straightforward:

$$J^{-1} = \frac{1}{3}\left( \hat{\xi}^{i}_{\vec{r}} \cdot \vec{r}_{\xi^i} \right) \quad or \tag{15.a}$$

$$J^{-1} = \frac{1}{3}\left( \hat{\xi}^{i}_{\vec{r}} \cdot \vec{r}_{\xi^i} \right)_{\xi^i}, \tag{15.b}$$

which is the same as that proposed by Abe et al. [21].

## 3. Approaches for arbitrary linear upwind schemes to achieve free-stream preservation

### 3.1. More discussions on free-stream preservation

In some literature [2], the analysis on *FSP* was based on Eq. (5) and started from Eq. (6), while the metric identity was checked thereafter numerically and/or theoretically. The acquisition of Eq. (5) relies on the presumption that constant fluxes can be moved outside of $\delta$. The process seems to be apparent at the first look, but less distinct appears when flux splitting is imposed. More discussions are given next, and only linear difference scheme is considered for simplicity.

Consider flux splitting of $\hat{E}$ at $\xi$ direction as $\hat{E} = \hat{E}^{+} + \hat{E}^{-}$. Suppose a $r$-th order scheme $\delta^{+}_{\xi}$ for $\hat{E}^{+}$ takes the form in Eq. (12) with $m_1 \geq n_1$, and $m_1+n_1 \geq r$. It is obvious that the symmetric counterpart for $\hat{E}^{-}$ will be $\delta^{-}_{\xi}\left(\cdot\right)_{i,j,k} = \frac{1}{\Delta}\sum_{i_1=-n_1}^{m_1} -a^{\xi}_{-i_1}\left(\cdot\right)_{i+i_1,j,k}$. If $m_1=n_1$ and $a^{\xi}_{i_1} = -a^{\xi}_{-i_1}$, the central scheme will be obtained and denoted as $\delta^{c}_{\xi}$. It is trivial that

$$\delta^{c}_{\xi}\hat{E}_{i,j,k}{}^{+} + \delta^{c}_{\xi}\hat{E}_{i,j,k}{}^{-} = \sum_{i_1=-m_1}^{m_1} a^{\xi}_{i_1}\left(\hat{E}^{+}_{i+i_1,j,k} + \hat{E}^{-}_{i+i_1,j,k}\right) = \sum_{i_1=-m_1}^{m_1} a^{\xi}_{i_1}\hat{E}_{i+i_1,j,k} = \delta^{c}_{\xi}\hat{E}_{i,j,k}. \tag{16}$$

So when the flow is uniform,

$$\begin{aligned}
&\delta^{c}_{\xi}\left(\hat{E}^{+}_{\infty} + \hat{E}^{-}_{\infty}\right) + \delta^{c}_{\eta}\left(\hat{F}^{+}_{\infty} + \hat{F}^{-}_{\infty}\right) + \delta^{c}_{\zeta}\left(\hat{G}^{+}_{\infty} + \hat{G}^{-}_{\infty}\right) \\
&= \delta^{c}_{\xi}\left(\hat{\xi}_{x}E_{\infty} + \hat{\xi}_{y}F_{\infty} + \hat{\xi}_{z}G_{\infty}\right) + \ldots \\
&= \left(\delta^{c}_{\xi}\hat{\xi}_{x} + \delta^{c}_{\eta}\hat{\eta}_{x} + \delta^{c}_{\zeta}\hat{\zeta}_{x}\right)E_{\infty} + \ldots \\
&= E_{\infty}\cdot\vec{I}_{x}{}^{*} + F_{\infty}\cdot\vec{I}_{y}{}^{*} + G_{\infty}\cdot\vec{I}_{z}{}^{*}
\end{aligned}$$

where $\vec{I}^{*}_{x_i} = \delta^{c}_{\xi}\hat{\xi}_{x_i} + \delta^{c}_{\eta}\hat{\eta}_{x_i} + \delta^{c}_{\zeta}\hat{\zeta}_{x_i}$ represents the numerical approximation of $\vec{I}_{x_i}$ as before. Through the above procedure, Eq. (6) is reached and *FSP* will be achieved if *MI* is established.

For the upwind scheme, $m_1=n_1$ and $a^{\xi}_{i_1} = -a^{\xi}_{-i_1}$ cannot be both fulfilled. Therefore $\delta^{+}_{\xi}\hat{E}^{+} + \delta^{-}_{\xi}\hat{E}^{-}$ cannot be re-arranged into a combinations of $\hat{E}$ like Eq. (16), and the constant fluxes would be difficult to be shift out of the difference operator. Consequently Eq. (6) and therefore *FSP* are thought to be hard to achieve by literatures [2, 9, 12], although the separate use of $\delta^{+}$, $\delta^{-}$ or $\delta^{c}$ on metrics and fluxes without splitting can yield *MI*. This difficulty will be addressed and solved in the next section.

3.2 Approaches to achieve free-stream preservation

In this section, it will be shown that through proper decomposition and careful concerns about flux splitting, *FSP* can be attained for arbitrary upwind schemes.

(1) Central scheme decomposition (CSD) of an upwind scheme

Consider the *r*-th order upwind scheme $\delta^{+}$ for the first-order derivative at position *i*, where indices like *j*, *k* are dropped for clarity. By Taylor expansion, there exists

$$\left(\delta^{+} f\right)_i = \partial f_i + a_{r+1}\partial^{(r+1)} f_i \times \Delta^{r} + a_{r+2}\partial^{(r+2)} f_i \times \Delta^{r+1} + \ldots, \tag{17}$$

where $a_i$ denotes the coefficient corresponding to the *i*-th order derivative.

Considering the symmetric property, it is easy to derive that for the counterpart $\delta_j^{-}$, when *r* is an even number, there will be

$$\left(\delta^{-} f\right)_i = \partial f_i + a_{r+1}\partial^{(r+1)} f_i \times \Delta^{r} - a_{r+2}\partial^{(r+2)} f_i \times \Delta^{r+1} + \ldots; \tag{18}$$

while if *r* is an odd number,

$$\left(\delta^{-} f\right)_i = \partial f_i - a_{r+1}\partial^{(r+1)} f_i \times \Delta^{r} + a_{r+2}\partial^{(r+2)} f_i \times \Delta^{r+1} + \ldots. \tag{19}$$

The sum of $\delta^{+}$ and $\delta^{-}$ can be uniformly expressed as,

$$\begin{aligned}\left[\left(\delta^{+} f\right)_i + \left(\delta^{-} f\right)_i\right]\Big/2 = \partial f_j + a_{2\lceil r/2\rceil+1}\partial^{(2\lceil r/2\rceil+1)} f_i \times \Delta^{2\lceil r/2\rceil} + \\ a_{2\lceil r/2\rceil+3}\partial^{(2\lceil r/2\rceil+3)} f_i \times \Delta^{2\lceil r/2\rceil+2} + \ldots\end{aligned}. \tag{20}$$

Eq. (20) shows that the summation represents a central scheme with the accurate order between *r* and the optimal order the combined stencils can afford.

Based on above understanding, a central operator $\delta^{c,(1)}$ is proposed as

$$\delta^{c,(1)} = \left[\left(\delta^{+} f\right)_i + \left(\delta^{-} f\right)_i\right]\Big/2, \tag{21}$$

where the number in superscript especially denotes the order of the derivative to approximate, namely, $\partial f$ here. Considering Eqs. (17) and (20),

$$\begin{aligned}\left(\delta^{c,(1)} f\right)_i - \left(\delta^{+} f\right)_i = -a_{2\lfloor r/2\rfloor+2}\partial^{(2\lfloor r/2\rfloor+2)} f_i \times \Delta^{2\lfloor r/2\rfloor+1} \\ - a_{2\lfloor r/2\rfloor+4}\partial^{(2\lfloor r/2\rfloor+4)} f_i \times \Delta^{2\lfloor r/2\rfloor+3} + \ldots,\end{aligned} \tag{22}$$

where only derivatives with even numbers exist on the right-hand side. Eq. (22) indicates its left-hand side regards a central discretization of $\partial^{(2\lfloor r/2\rfloor+2)} f_i$. Define $\delta^{c,(2\lfloor r/2\rfloor+2)}$ as

$$\delta^{c,(2\lfloor r/2\rfloor+2)} = \left(\delta^{c,(1)} f\right)_i - \left(\delta^{+} f\right)_i. \tag{23}$$

It can be conceived that the expansion of Eq. (23) will have the form $\frac{1}{\Delta}\sum_{i_1=-m'}^{m'} a_{i_1}\left(\bullet\right)_{i+i_1}$ with

$a_{i_1} = -a_{-i_1}$, $m'$=max($m_1$, $n_1$), and

$$\sum_{i_1=-m'}^{m'} a_{i_1} = 0 \tag{24}$$

Similar analysis can be made toward $\delta_j^-$, and the following decompositions are obtained:

$$\begin{cases} \delta^+\left(f\right)_i = \delta^{c,(1)}\left(f\right)_i - \delta_c^{c,(2\lfloor r/2\rfloor+2)}\left(f\right)_i \\ \delta^-\left(f\right)_i = \delta^{c,(1)}\left(f\right)_i + \delta^{c,(2\lfloor r/2\rfloor+2)}\left(f\right)_i \end{cases}. \tag{25}$$

Because $\delta^{c,(1)}$ and $\delta^{c,(2\lfloor r/2\rfloor+2)}$ are both certain central schemes, the decomposition is referred as central scheme decomposition or CSD. Next, CSD of three upwind schemes are presented as examples.

(a) CSD of the linear fifth-order WENO scheme.

The case represents an example for schemes with odd order numbers. For $\delta^+$:

$$\delta^+\left(f\right)_i = \frac{1}{60\Delta}\left(-2f_{i-3} + 15f_{i-2} - 60f_{i-1} + 20f_{\ i} + 30f_{i+1} - 3f_{i+2}\right), \tag{26}$$

then accordingly,

$$\delta^-\left(f\right)_j = \frac{-1}{60\Delta}\left(-3f_{i-2} + 30f_{i-1} + 20f_i - 60f_{\ i+1} + 15f_{i+2} - 2f_{i+3}\right).$$

Following above procedures, $\delta^{c,(1)}$, $\delta^{c,(6)}$ and their Taylor expansion can be derived and shown in Table 1.

Table 1. CSD of the linear fifth-order WENO scheme and corresponding Taylor expansions

| Operator | Form | Taylor expansion |
|---|---|---|
| $\delta^{c,(1)}$ | $\frac{1}{60\Delta}\begin{pmatrix} -f_{i-3} + 9f_{i-2} - 45f_{i-1} + \\ 45f_{i+1} - 9f_{i+2} + f_{i+3} \end{pmatrix}$ | $f_i' + \frac{1}{140}f_i^{(7)}\Delta^6 + \frac{1}{720}f_i^{(9)}\Delta^8 + \dots$ |
| $\delta^{c,(6)}$ | $\frac{1}{60\Delta}\begin{pmatrix} f_{i-3} - 6f_{i-2} + 15f_{i-1} - 20f_i + \\ 15f_{i+1} - 6f_{i+2} + f_{i+3} \end{pmatrix}$ | $\frac{1}{60}f_i^{(6)}\Delta^5 + \frac{1}{240}f_i^{(8)}\Delta^7 + \dots$ |

(b) CSD of the second-order upwind scheme

The case shows an example for schemes with even order numbers. The form of $\delta^{+}$ is:

$$\delta^{+}=\frac{1}{2\Delta}\left(f_{i-2}-4f_{i-1}+3f_{i}\right). \tag{27}$$

Similarly, $\delta^{c,(1)}$ and $\delta^{c,(4)}$ and their Taylor expansions can be summarized in Table 2. It is worth mentioning that $\delta^{c,(1)}$ is a second-order discretization and not of the optimal fourth-order the dependent stencil can support.

Table 2. CSD of the linear second-order upwind scheme and corresponding Taylor expansions

| Operator | Form | Taylor expansion |
|---|---|---|
| $\delta^{c,(1)}$ | $\frac{1}{4\Delta}\left(f_{i-2}-4f_{i-1}+4f_{i+1}-f_{i+2}\right)$ | $f_i'-\frac{1}{3}f_i^{(3)}\Delta^2-\frac{7}{60}f_i^{(5)}\Delta^4+\ldots$ |
| $\delta^{c,(4)}$ | $\frac{1}{4\Delta}\left(-f_{i-2}+4f_{i-1}-6f_i+4f_{i+1}-f_{i+2}\right)$ | $-\frac{1}{4}f_i^{(4)}\Delta^3-\frac{1}{24}f_i^{(6)}\Delta^5+\ldots$ |

(c) CSD of a third-order mixed node/half-node scheme

In Ref. [22], H. X. Zhang proposed a method to derive the high-order conservative schemes. Following the idea, we recently derived series of high-order mixed node/half-node schemes. The linear form of the third-order case is:

$$\delta^{+}(f)_i=\frac{1}{\Delta}\times\left[\frac{1}{3}\left(f_i-f_{i-1}\right)+\frac{1}{6}\left(5f_{i+1/2}-6f_{i-1/2}+f_{i-3/2}\right)\right]. \tag{28}$$

The advantage of this formed scheme lies in its small grid stencil, which is favorable in the realization of the nonlinear counterpart. More details about this regard will be discussed in other publications.

Similarly, $\delta^{c,(1)}$, $\delta^{c,(4)}$ and their Taylor expansion are summarized in Table 3.

Table 3. CSD of the linear third-order upwind scheme and corresponding Taylor expansions

| Operator | Form | Taylor expansion |
|---|---|---|
| $\delta^{c,(1)}$ | $\frac{1}{12\Delta}\left[\begin{array}{l}2\left(f_{i+1}-f_{i-1}\right)+\\ \left(-f_{i+3/2}+11f_{i+1/2}-11f_{i-1/2}+f_{i-3/2}\right)\end{array}\right]$ | $f_i'-\frac{7}{960}f_i^{(5)}\Delta^4-\frac{f_i^{(7)}\Delta^6}{2016}+\ldots$ |
| $\delta^{c,(4)}$ | $\frac{1}{12\Delta}\left[\begin{array}{l}2\left(f_{i+1}-2f_i+f_{i-1}\right)+\\ \left(-f_{i+3/2}+f_{i+1/2}+f_{i-1/2}-f_{i-3/2}\right)\end{array}\right]$ | $-\frac{1}{48}f_i^{(4)}\Delta^3-\frac{5}{2304}f_i^{(6)}\Delta^5+\ldots$ |

In summary, through CSD, arbitrary upwind schemes can be decomposed into two central schemes.

(2) Requirements caused by flux splitting to achieve free-stream preservation

Because $\delta^{+}$ and $\delta^{-}$ act on different split fluxes, it seems that in Eq. (25), only $\delta^{\mathrm{c},\,(1)}$ satisfies *FSP* according to former discussions. Next, it will be shown that *FSP* can be fulfilled through Eq. (25) if proper flux splitting method is used.

Inspired by Lax-Friedrichs splitting method, let's consider a scheme as

$$\hat{E}^{\pm}=\frac{1}{2}\left(\hat{E}\pm\hat{A}\cdot Q\right)\ \text{ or }\ \hat{E}^{\pm}=\frac{1}{2}\left(\hat{E}\pm\hat{E}_{ref}\right),\tag{29}$$

where $\hat{A}$ denotes certain constant matrix or number and $\hat{E}_{ref}$ represents some referenced flux. Especially, when the uniformed-flow condition is imposed, $\hat{A}$ and $\hat{E}_{ref}$ would be locally constant at least at the dependent stencil of $\delta^{c,(2\lceil r/2\rceil+2)}$. Then by Eqs. (24) and (29),

$$\delta^{c,(2\lceil r/2\rceil+2)}\left(\hat{E}_{\infty}^{+}\right)\text{-}\delta^{c,(2\lceil r/2\rceil+2)}\left(\hat{E}_{\infty}^{-}\right)=\begin{cases}\dfrac{1}{2}\sum\limits_{i_1=-m'}^{m'}a_{i_1}\left(\hat{E}_{i+i_1}-\hat{E}_{i+i_1}\right)_{\infty}+\dfrac{\hat{A}Q_{\infty}}{\Delta}\sum\limits_{i_1=-m'}^{m'}a_{i_1}=0 \quad or\\ \dfrac{1}{2}\sum\limits_{i_1=-m'}^{m'}a_{i_1}\left(\hat{E}_{i+i_1}-\hat{E}_{i+i_1}\right)_{\infty}+\dfrac{\hat{E}_{ref}}{\Delta}\sum\limits_{i_1=-m'}^{m'}a_{i_1}=0\end{cases}.\tag{30}$$

So because of the zero contributions of $\delta^{c,(2\lceil r/2\rceil+2)}$ in $\delta^{+}\hat{E}_{\infty}^{+}+\delta^{-}\hat{E}_{\infty}^{-}$, only the action of $\delta^{\mathrm{c},\,(1)}$ left under the uniform flow. Hence *FSP* is fulfilled for arbitrary upwind schemes through CSD.

In the following, a Lax-Friedrichs-type scheme is given as an example. Consider the original splitting form

$$\hat{E}^{\pm}=\frac{1}{2}\left(\hat{E}\pm\Lambda\cdot Q\right),\tag{31}$$

where $\Lambda\text{=}diag(\lambda_1,...,\lambda_5)=\left(\hat{U},\hat{U},\hat{U},\hat{U}-c\left|\hat{\xi}\right|,\hat{U}+c\left|\hat{\xi}\right|\right)$, $\hat{U}=\hat{\xi}_{x_j}u^{j}$, $c$ is the sound speed and $\left|\hat{\xi}\right|=\left(\hat{\xi}_{x_j}\cdot\hat{\xi}_{x_j}\right)^{1/2}$. Let $\hat{A}$ in Eq. (29) as $diag\left(\max\lambda_1,...,\max\lambda_5\right)$, where the maximum value of $\lambda_{\mathrm{i}}$ should be obtained over the whole field or the dependent stencil of $\delta^{\mathrm{c},\,(1)}$ from $i\text{-}m'$ to $i\text{+}m'$, then Eq. (30) will be established. In this study the whole field is chosen for simplicity.

In short, for arbitrary linear upwind scheme to approximate the flux derivative in each coordinate, if $\delta^{c,(1)}$ by CSD is used for metric approximation and aforementioned flux splitting is adopted, *FSP* can be achieved.

(3) Directionally consistent interpolation for half-node or mixed type schemes

It is straightforward that above algorithms work for node-type difference schemes. While for half-node or mixed schemes, extra considerations should be cared. To evaluate the flux $\hat{E}^{\pm}$ at half nodes, it is trivial that metrics should be available at the same locations as well. There are at least two ways of acquisition:

(a) Derive the geometric information at half nodes first which include the coordinates and their derivatives, and then use $\delta^{c,(1)}$ for metric evaluation. For the coordinates at half nodes, the linear interpolation is used; for their derivatives, the same $\delta^{c,(1)}$ is used again and coordinates needed at half nodes are attained by interpolation once more.

(b) Evaluate the metrics at nodes by $\delta^{c,(1)}$ first, while coordinates and their derivatives at half nodes are still needed. At half nodes, coordinates can be obtained by interpolation, and their derivatives are evaluated by telescoping the $\delta^{c,(1)}$ to the half nodes. Similarly, the acquisition of derivatives might need coordinates at half nodes once again, which will be obtained by interpolation from coordinates at nodes. After the metrics at nodes are available, they are interpolated to the half nodes at last [12]. In Ref. [12], the sixth- or fourth-order interpolation was suggested as candidate.

Considering the interpolation and the difference being commutable, it is supposed the two implementations are equivalent, and aforementioned second implementation seems easier for numerical realization and therefore is used in this study. To further clarify the relationship between the interpolation and *MI*, four cases are designed to investigate for $I_{\vec{r}}$ on a 3-D randomized grid in Section 4.1 with the dimension $41^3$. In the study, $\delta^{c,(1)}$ in Table 3 is used and three interpolations are chosen with fourth to sixth order [23-24] shown in Table 4.

Table 4. Forms of fourth-, fifth- and sixth-order interpolations

| Order | Forms of interpolations |
|---|---|
| 4 | $f_{i+1/2} \approx \frac{1}{16}\left(-f_{i-1} + 9f_i + 9f_{i+1} - f_{i+1}\right)$ |
| 5 | $f_{i+1/2} \approx \frac{1}{128}\left(3f_{i-2} - 20f_{i-1} + 90f_i + 60f_{i+1} - 5f_{i+2}\right)$ |
| 6 | $f_{i+1/2} \approx \frac{1}{256}\left(3f_{i-2} - 25f_{i-1} + 150f_i + 150f_{i+1} - 25f_{i+2} + 3f_{i+3}\right)$ |

Four cases are considered:

Case I: In the evaluation of $\hat{\xi}^i_{\vec{r}}$ and then $I_{\vec{r}}$, the fourth-order interpolation is uniformly used in three curvilinear coordinate directions.

Case II: In the evaluation of $\hat{\xi}^i_{\vec{r}}$, the same interpolations are chosen as in Case I. In the computation of $I_{\vec{r}}$ afterwards, the sixth-order interpolation is used for $\eta$ direction and fourth-order one is used for rest directions.

Case III: In the evaluation of $\hat{\xi}^i_{\vec{r}}$, the fourth-order interpolation is used for $\xi$ direction, the fifth-order one is used for $\eta$ direction, and the sixth-order one is used for $\zeta$ direction. Afterwards, the same choices of interpolation are used for the evaluation of $I_{\vec{r}}$ .

Case IV: In the evaluation of $\hat{\xi}^i_{\vec{r}}$, the same interpolations are chosen as in case III. Afterwards in the computation of $I_{\vec{r}}$ , only the sixth-order interpolation is used for $\xi$ direction, while the rest choices of interpolations are the same as that in case III.

In the above four situations, the value of $I_{\vec{r}}$ is computed and shown in Table 5, where $\|\cdot\| = \sqrt{\frac{\sum_{i=1}^{N}(\cdot)_i^2}{N}}$ and *N* is the total grid number.

Table 5. Values of $I_{\vec{r}}$ in four cases of the implementation of interpolations

| Cases | $\| I_x \|$ | $\| I_y \|$ | $\| I_z \|$ |
|---|---|---|---|
| Case I | 5.875271E-012 | 5.846736E-012 | 5.786077E-012 |
| Case II | 23.911432 | 33.977481 | 24.252242 |
| Case III | 6.25142E-012 | 6.373999E-012 | 6.570719E-012 |
| Case IV | 40.855302 | 29.04189 | 28.667106 |

It is obvious that case II and IV violate *MI*. Based on numerical experiment, we propose an idea of directionally consistent interpolation (namely DCI) as: in order to achieve *MI* for half-node or mixed schemes, the consistent linear interpolation should be imposed on each coordinate direction in the evaluation of metrics and fluxes, while the interpolations could be different in different directions. Further analytic proof is undergoing and will come into sight soon.

3.3 Short summary on numerical implementations

Based on the above analysis, a summary will be made regarding numerical implementations:

(1) Eq. (7) or (8) are chosen as the forms of grid metrics, and the latter is used in this study.

(2) Given any linear upwind schemes $\delta^+$ (and $\delta^-$ accordingly), $\delta^{c,(1)}$ is derived by Eq. (21) to compute the metrics. Particularly, flux splitting method described in Section 3.2 should be used.

(3) For half-node or mixed schemes, interpolations must be used in above steps, where DCI should be followed to achieve *MI* and therefore *FSP*.

(4) Eq. (15) is suggested for the derivation of Jacobian, where Eq. (15.b) is chosen in this study. Especially, $\hat{\xi}^i_{\vec{r}}$ in the equation should be evaluated by step (1) - (3).

(5) For fluxes in Eq. (5), the given $\delta^{\pm}$ is used for approximations.

The above procedures is valid for central schemes as well except that the restriction for flux splitting can be released, e.g., the fourth-order central scheme can be both used for the flux and metric evaluation. Moreover, the idea of the analysis can also be applied to the construction of conservative schemes, i.e., for $h_{i+1/2}$ in $(\partial f/\partial x)_i=(h_{i+1/2}-h_{i-1/2})/\Delta x$. More details will be discussed in other publications.

## 4. Numerical validations

In this section, two canonical problems are tested by using 2-D Euler equations, i.e., one regarding *FSP* and the other about the isentropic vortex preservation, which are favored by studies on metric-induced errors. Two upwind schemes for spatial discretizations are used, namely, the fifth-order upwind scheme by Eq. (26) (UPW5) and the third-order mixed upwind scheme by Eq. (28) (M-UPW3). To combine with M-UPW3, the fourth-order interpolation in Table 5 is used to derive variables at half nodes. For reference, the fourth-order central scheme (CS4) is also realized. To enhance its numerical stability in some computations, a sixth-order compact filter (CF6) [23] is used as:

$$\alpha_f \bar{f}_{i-1}+\bar{f}_i+\alpha_f \bar{f}_{i+1}=\frac{1}{2}\begin{bmatrix}\left(\frac{11}{16}+\frac{5\alpha_f}{8}\right) f_i+\left(\frac{15}{32}+\frac{17\alpha_f}{16}\right)\left(f_{i+1}+f_{i-1}\right)+ \\ \left(\frac{-3}{16}+\frac{3\alpha_f}{8}\right)\left(f_{i+2}+f_{i-2}\right)+\left(\frac{1}{32}-\frac{\alpha_f}{16}\right)\left(f_{i+3}+f_{i-3}\right)\end{bmatrix},$$

where $\alpha_f$=0.45 in this study. Other details are explained in Section 3.3. For temporal algorithm, the third-order TVD Rung-Kutta method is used [24].

To provide a tough test, three nonuniform grids are chosen including two seriously deformed ones. Their generations are explained first.

### 4.1 Grid configurations

The grids are of three categories: wavy grids, randomized grids and triangular grids.

(1) Wavy grids [20]

The grid coordinates are generated by:

$$\begin{cases} x_{i,j}=-\dfrac{L}{2}+\dfrac{L}{I_{max}-1}\left[(i-1)+A_x\sin\dfrac{n_{xy}\pi(j-1)}{J_{max}-1}\right] \\ y_{i,j}=-\dfrac{L}{2}+\dfrac{L}{J_{max}-1}\left[(j-1)+A_y\sin\dfrac{n_{xy}\pi(i-1)}{I_{max}-1}\right]\end{cases},$$

where $L$=16, $i$=1...$I_{max}$, $j$=1..., $J_{max}$, $L$=16, $A_x$ =0.4($I_{max}$-1)/$L$, $A_y$ =0.8($J_{max}$-1)/$L$, and $n_{xy}$=6. Two sets of grid number for ($I_{max}\times J_{max}$) are chosen as: (41×41) and (81×81). The grids with the number (41×41) is shown in Fig. 1.

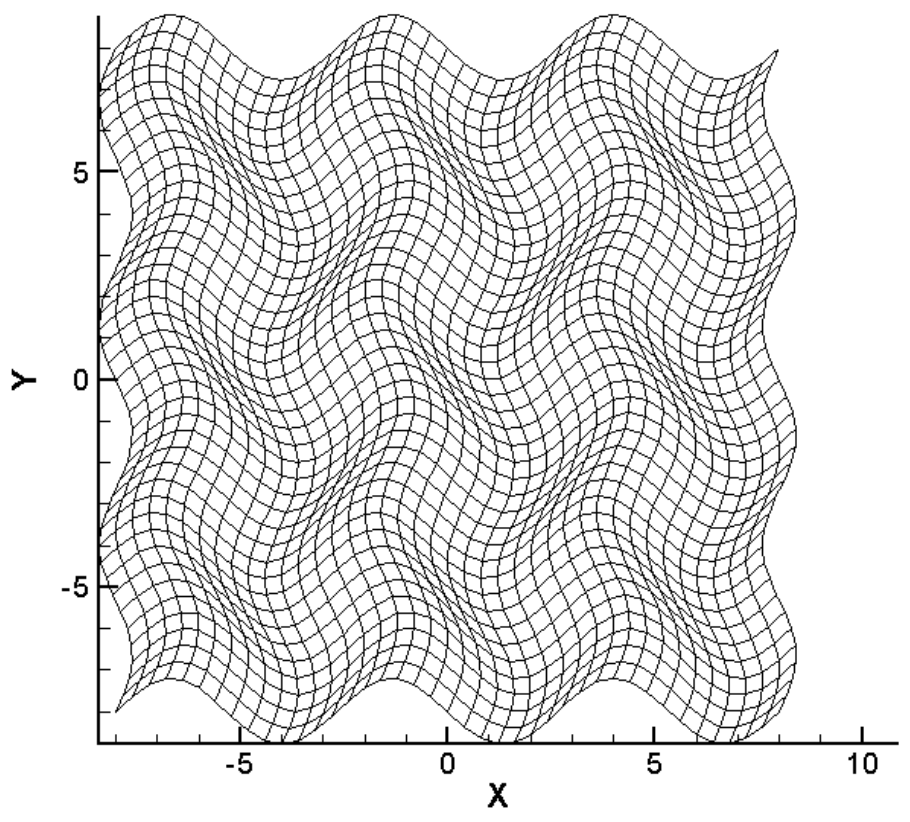

Fig. 1. Wavy grid with the number 41×41

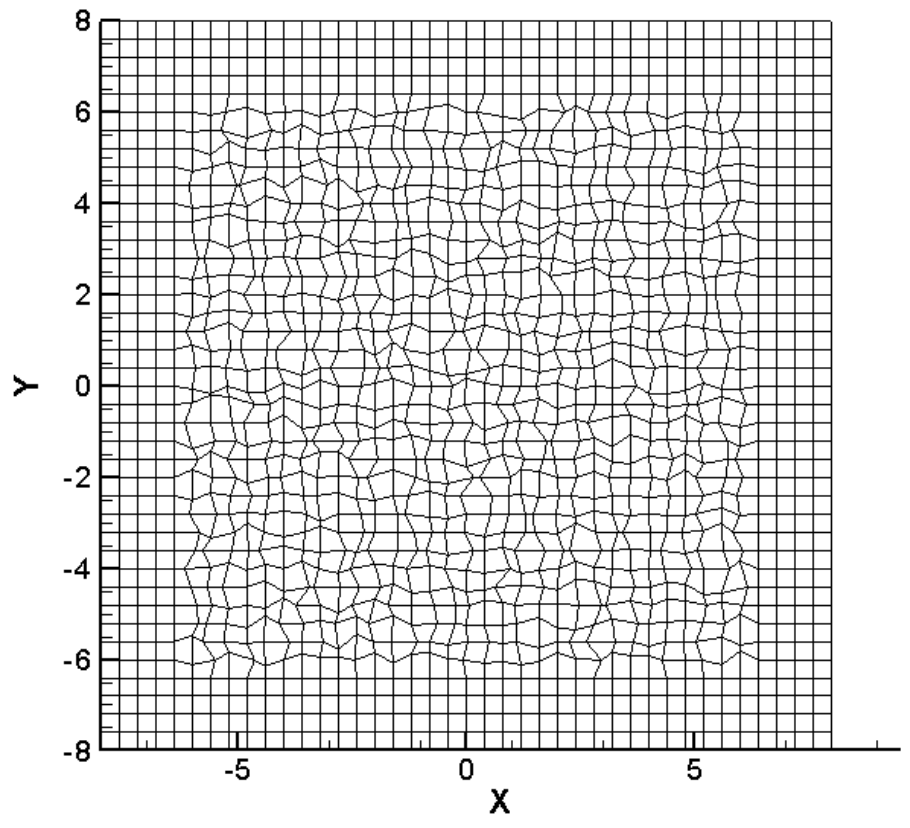

Fig. 2. Randomized grid with the number 41×41

(2) Randomized grids [1]

(a) 2-D case

The coordinates are generated by:

$$x_{i,j} = -\frac{L}{2} + \frac{L}{I_{\max} - 1}\Big[\big(i-1\big) + 2A_{i,j}\big(Rand(0,1) - 0.5\big)Rand(0\,|\,1)\Big]$$
$$y_{i,j} = -\frac{L}{2} + \frac{L}{J_{\max} - 1}\Big[\big(j-1\big) + 2A_{i,j}\big(Rand(0,1) - 0.5\big)(1 - Rand(0\,|\,1))\Big]$$

where $L$=16, $A_{ij}$ equals 0.45 at $i$=5...$I_{\max}$-4 or $j$=5...$J_{\max}$-4 otherwise equals zero, Rand(0, 1) is a random function ranging from 0 to 1 while Rand(0|1) is one having the value 0 or 1. Two sets of grid number are chosen as (41×41) and (81×81). The grids with the number (41×41) is shown in Fig. 2. It is worthy to mention the randomized grid here has the largest deformation by $A_{ij}$ =0.45 than that reported in previous literatures [1, 2, 20] with $A_{ij}$ =0.4, and further increase of $A_{ij}$ will cause negative grid-cell area.

(b) 3-D case

The grid generation is similar to that of 2-D case, which is still through randomizing uniform grids with 0.45 magnitude grid spacing in a random direction.

(3) Triangular grids

In order to explore the potential of the proposed methodology, a triangular grid is designed to mimic the unstructured grid. The construction of grids is illustrated in Fig. 3. In Fig. 3(a), a series of square cells are first built; then pairs of points collapse into one like ($A_2$, $A_3$)→$A_{2,3}$, while they are still treated as two separate points in the computation. The final grid looks like the one in Fig. 3(b), which resembles typical unstructured topology.

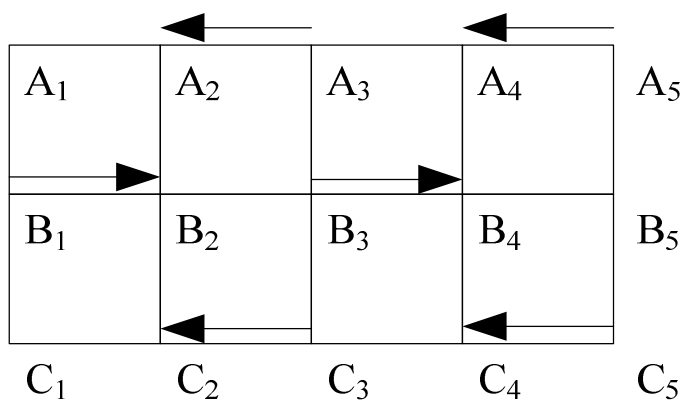

(a) Before

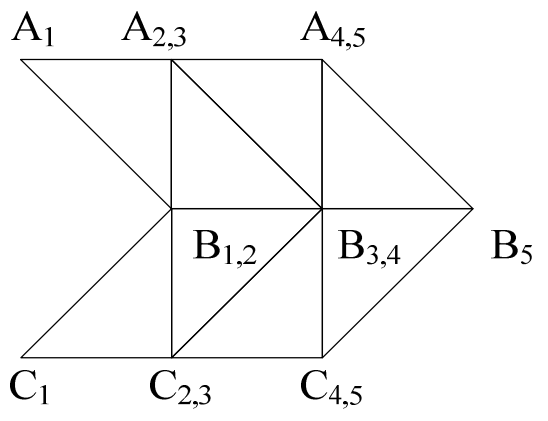

(b) After

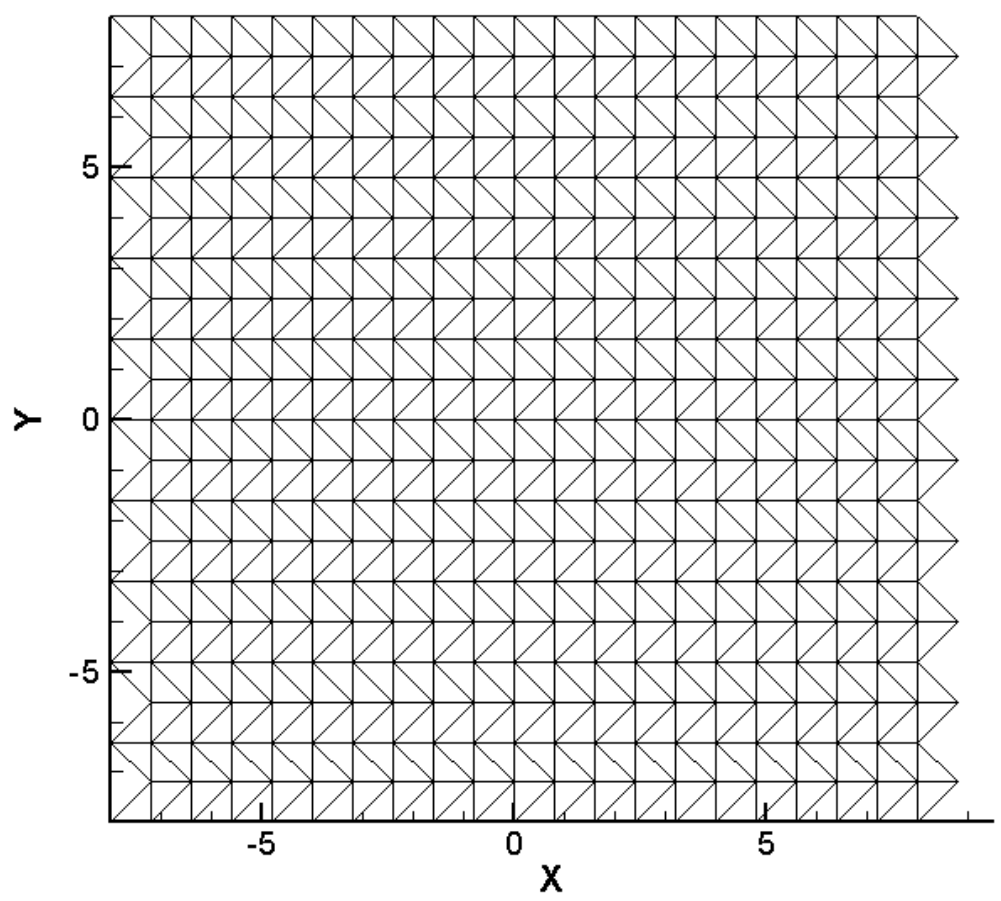

(c) Grid with the number 41×21

Fig. 3. Generation of the triangular grid

The computation field is [-8, 8]×[-8, 8], and two sets of grid number are chosen as (41×21) and (81×41). Fig. 3(c) shows the grid with the number (41×21).

In computations, the periodic boundary condition is employed for all cases, which is realized by extending extra four layers of grids on four sides, through which grid metrics and Jacobian can be calculated by integrally using the aforementioned methods.

4.2 Check on *FSP* on the randomized grid

This test is conducted on the 3-D randomized grid with the number $41^3$. Three schemes are used, i.e., CS4, M-UPW3 and UPW5. A free-stream condition is imposed with the Mach number as 1. The computation runs until $t$=10 with the time step $\Delta t$=0.01. $L_2$ errors of velocity component $v$ and $w$ are shown in Table. 6.

Table 6. $L_2$ errors of $v$ and $w$-component in *FSP* test on the randomized grid

| Scheme | $v$-component | $w$-component |
|---|---|---|
| CS4 | 5.040337010914540E-014 | 5.048442957102751E-014 |
| M-UPW3 | 5.450028935790510E-015 | 5.442990712969394E-015 |
| UPW5 | 2.421037216639132E-015 | 2.520031678789157E-015 |

It can be seen from Table 6 that the methodology proposed for upwind node and mixed type schemes are validated to achieve *FSP*. According to previous discussions, CS4 is expected to fulfill *FSP* as well, which is also verified by the computation.

4.3 Vortex preservation on three types of grids [1, 20]

This problem is rather popular to investigate the performance of numerical schemes on deformed grids. The flow is non-dimensionalized by the density and the speed of sound, and the free-stream Mach number is one. An isentropic vortex is initially superimposed on the uniformed flow at $\vec{r}_0$ =(0, 0) as [20]

$$\begin{cases} (\delta u, \delta v) = \varepsilon \tilde{r} e^{\alpha(1-\tilde{r}^2)} (\sin\theta, -\cos\theta) \\ \delta T = -\dfrac{(\tilde{r}-1)\varepsilon^2}{4\alpha\tilde{r}} e^{2\alpha(1-\tilde{r}^2)} \\ \delta S = \delta(p/\rho^\gamma) = 0 \end{cases},$$

where $\tilde{r} = |\vec{r} - \vec{r}_0| / r_c$, $r_c$=1, $\alpha$=0.204, $\varepsilon$=0.3, and $\gamma$=1.4.

The computation runs from above initial conditions for a time $t$=16 at $\Delta t$=0.01. The period corresponds to one movement circle of the vortex to return to its initial place through the periodic boundary. Three different types of meshes are chosen and different schemes are comparatively investigated.

(1) Wavy grids

Three sets of grid numbers are chosen: (41×41), (81×81) and (161×161). CS4, M-UPW3 and UPW5 are used in the computation. As the representative, contours of vorticity magnitude on the grid (41×41) is shown in Fig. 4(a)-(c), and the distribution of $v$-component along the line $j$=20 is depicted in Fig. 4(d). The pressure is not chosen for visualization because of its relatively smooth distribution. Although three methods achieve *MI* and *FSP* theoretically, the result of CS4 appears noisy for lacking dissipation. The quantitative check in Fig. 4(d) shows M-UPW3 and UPW5 demonstrate a reasonable description about the vortex profile, while M-UPW3 behaves more smearing; on the other hand, CS4 yields a result with oscillations with short wavelength at the smooth region away from the vortex. Hence the methodology developed for upwind schemes manifests its advantage over central schemes if the treatment is absent like filtering.

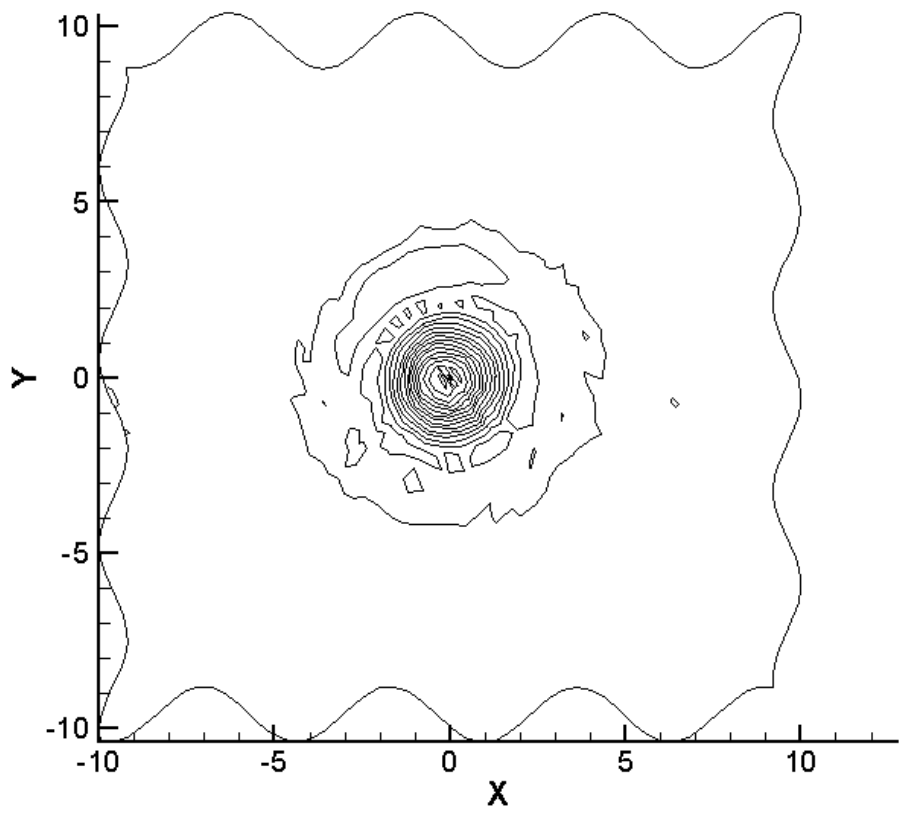

(a) CS4

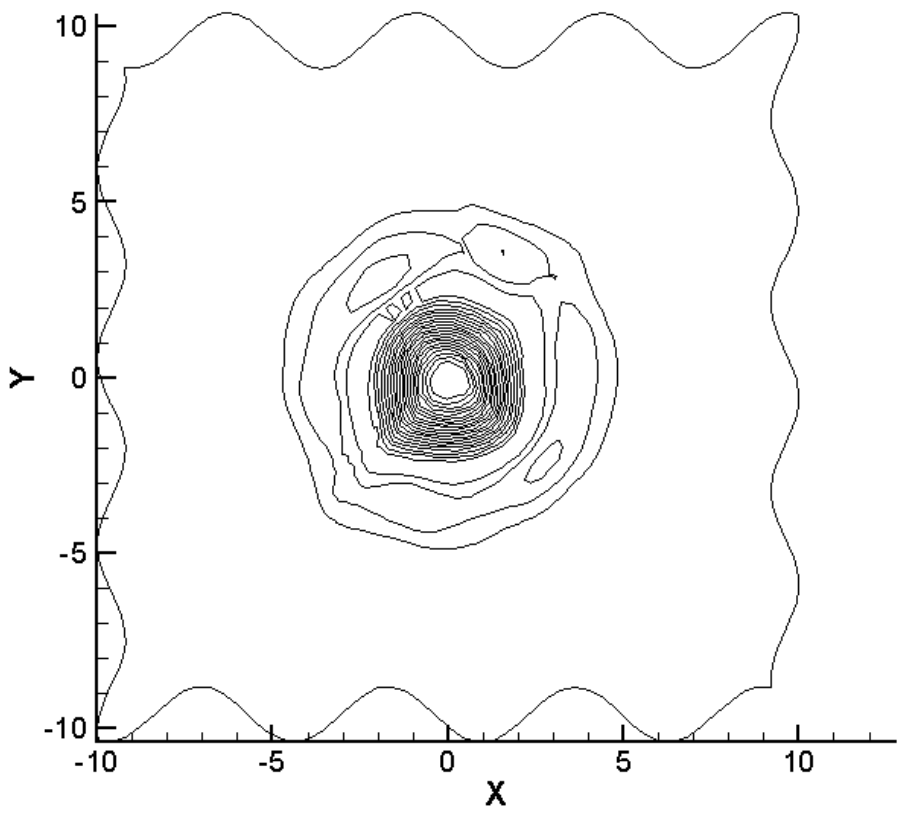

(b) M-UPW3

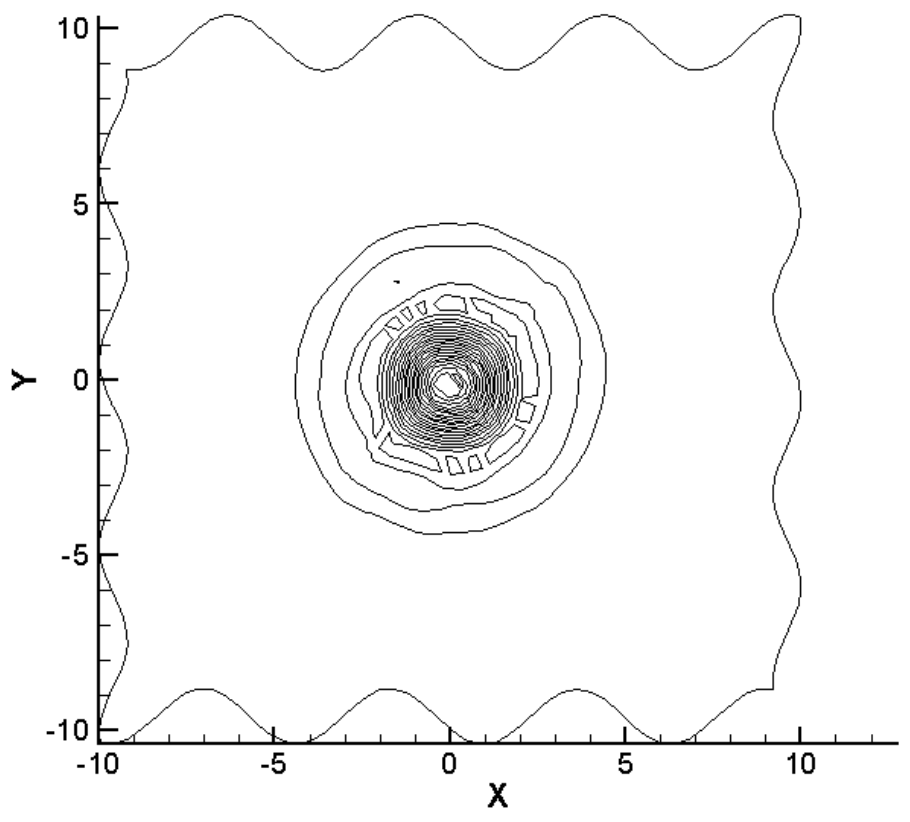

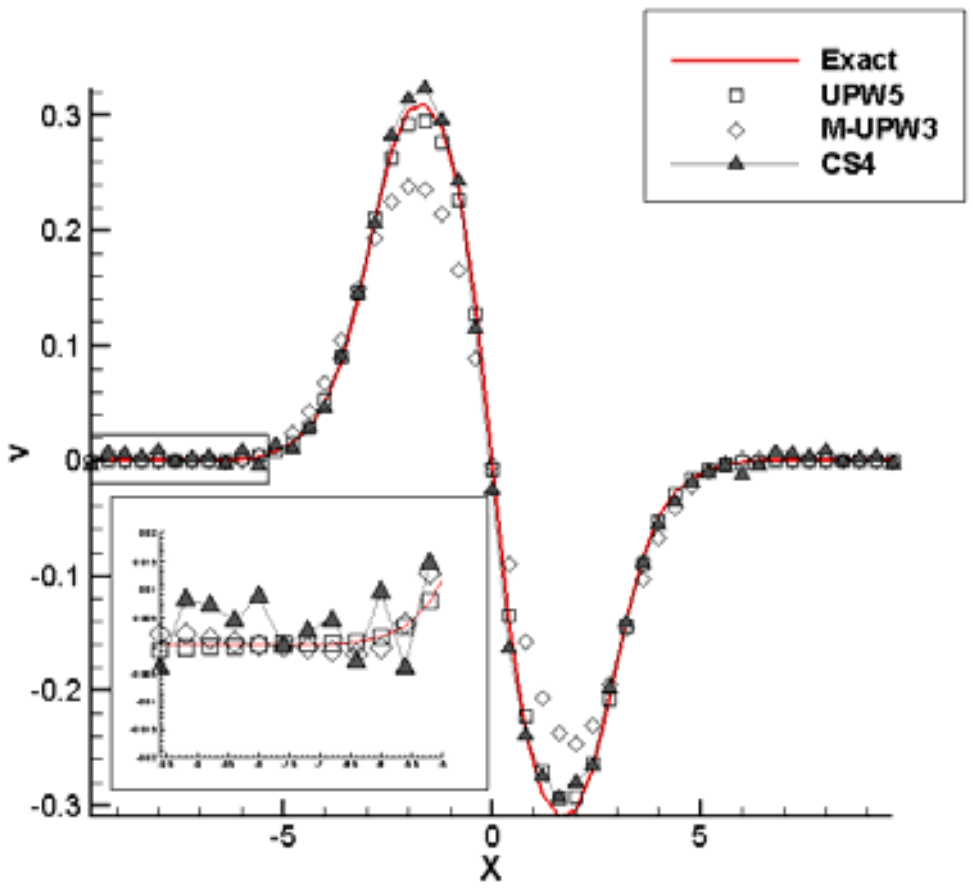

(c) UPW5

(d) Distributions of *v*-component

Fig. 4. Vorticity contours and v-distributions along the line with $j = J_{max}/2+1$ on wavy grids in moving vortex problem

Similar computations are made on the rest two grids and results with the convergence are obtained. Taking the use of computation errors, the accuracy orders of schemes can be derived and are shown in Table 7. As in Ref. [20], the order of schemes on the wavy grids are smaller than their analytic counterpart. The order of CS4 at the grid (81×81) unexpectedly has a large value, which might be caused by unsmooth process during the grid convergence.

Table 7. $L_2$ errors in the v-component in moving vortex problem on wavy grids

| Grids | CS4 | | M-UPW3 | | UPW5 | |
|---|---|---|---|---|---|---|
| | $L_2$ errors | order | $L_2$ errors | order | $L_2$ errors | order |
| 41×41 | 6.839435E-02 | — | 2.321206E-01 | — | 4.670484E-02 | — |
| 81×81 | 4.818256E-03 | 3.8273 | 6.411272E-02 | 1.85626 | 3.238539E-03 | 3.85074 |
| 161×161 | 4.213631E-04 | 3.5154 | 1.234671E-02 | 2.37648 | 1.558367E-04 | 4.37711 |

(2) Randomized grids

Computations are made on two grids with the number (41×41) and (81×81), where CS4+CF6, W-UPW3 and UPW5 are checked. In this situation, CS4 cannot work independently unless aforementioned sixth-order filter is used. Again, contours of vorticity magnitude on the grid (41×41) is shown in Fig. 5(a)-(c) and the distribution of *v*-component along the line $j = J_{max}/2+1$ is depicted in Fig. 5(d). On such seriously deformed grid, two upwind schemes indicate their robustness and fair performance on vortex preservation. Their solutions about *v*-component show rather smooth distributions as well, where M-UPW3 appears relatively more dissipative. With the help of filtering, CS4 works normally and generates a result comparable to that of UPW5.

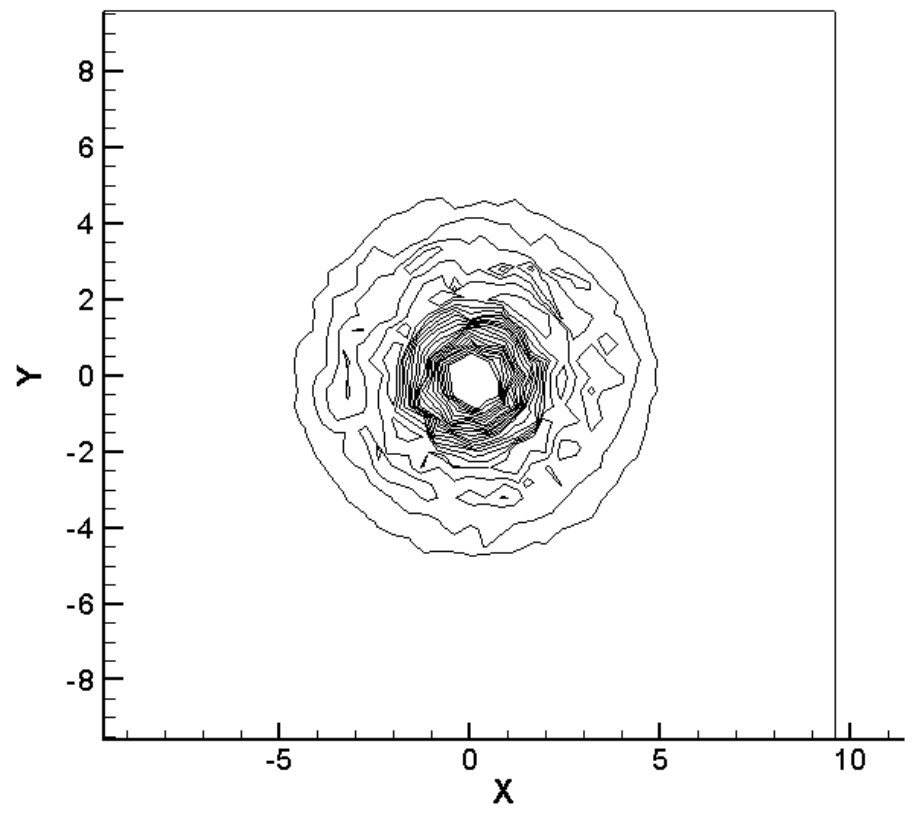

(a) CS4+CF6

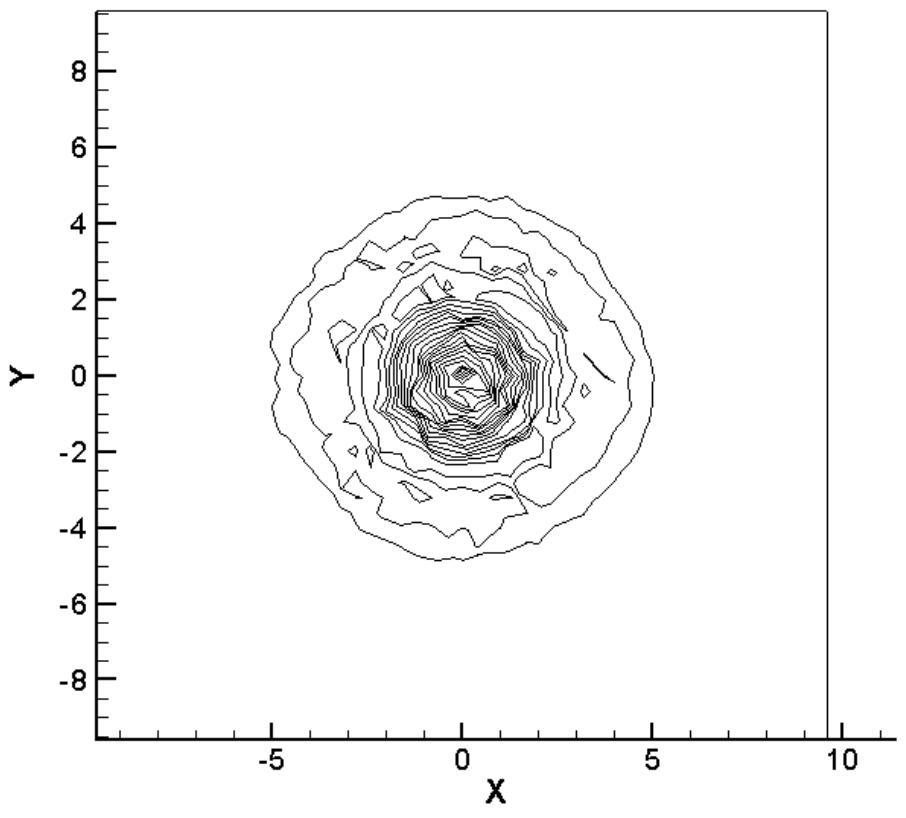

(b) M-UPW3

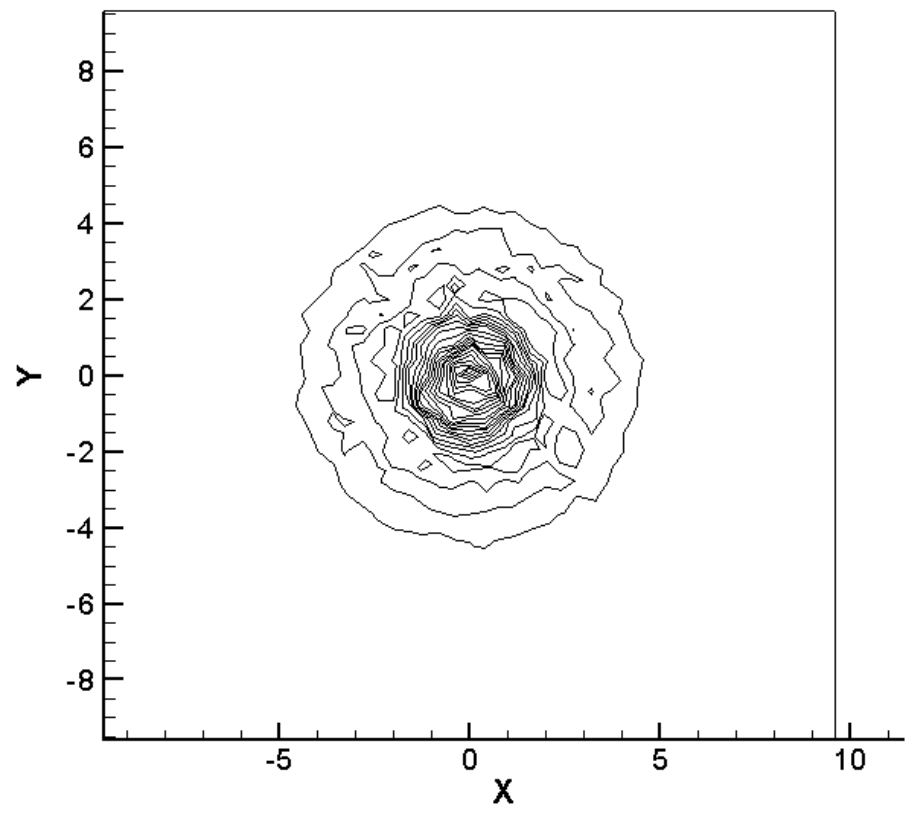

(c) UPW5

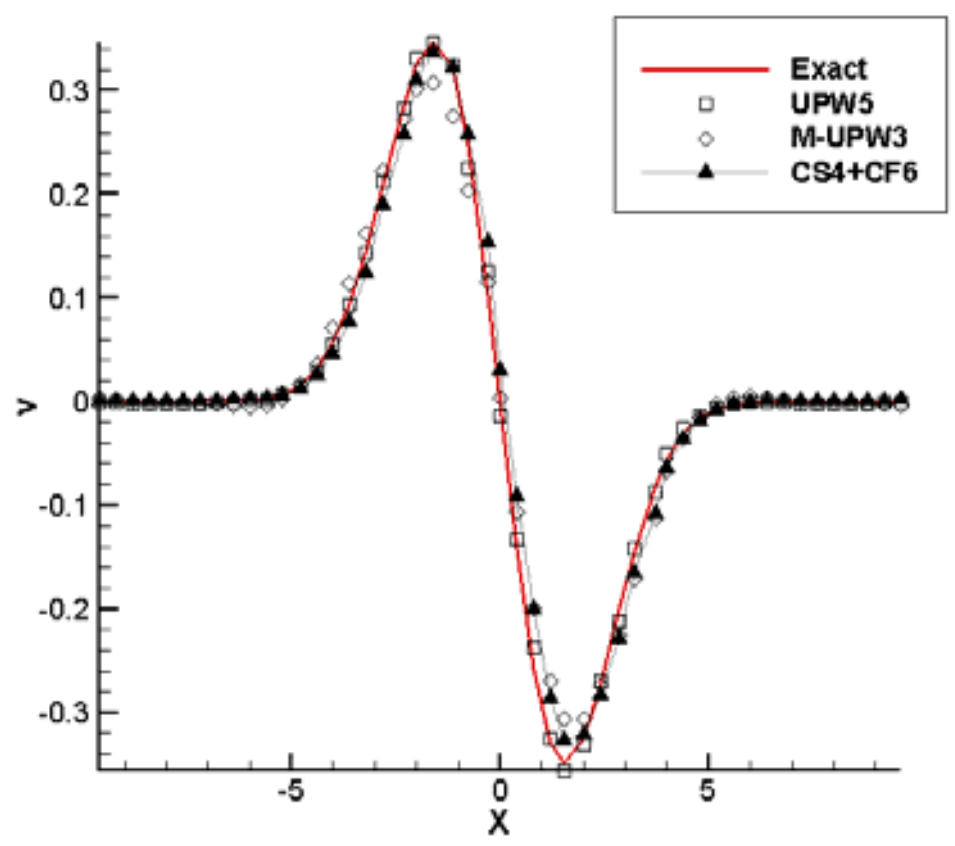

(d) Distributions of *v*-component

Fig. 5. Vorticity contours and v-distributions along the line $j= J_{max}/2+1$ on randomized grids in moving vortex problem

The results on the grid with number (81×81) are similar to that of the coarse grid, except for the decreased wavelength of irregularities in vorticity contours. They are omitted for visualization thereby.

(3) Triangular grids

This case provide a situation analogous to unstructured grid. Two grid numbers are set as (41×21) and (81×41) and three schemes are checked, namely, CS4+CF6, M-UPW3 and UPW5. The individual use of CS4 does not work once more. The vorticity contours on two grids are first shown in Fig. 6, which manifest the potential of difference schemes to solve problems on structured-like grids if *MI* is fulfilled. It is interesting to observe that on the coarse grid, the vorticity contour by CS4+CF6 appears asymmetric compared with that of M-UPW3 and UPW5. What is more, extra perturbations emerge near the upper and lower boundaries by the central scheme, while upwind schemes yield relative clean results. When the grid number is increased to (81×41), such difference becomes far from obvious because of the convergence to the exact solution.

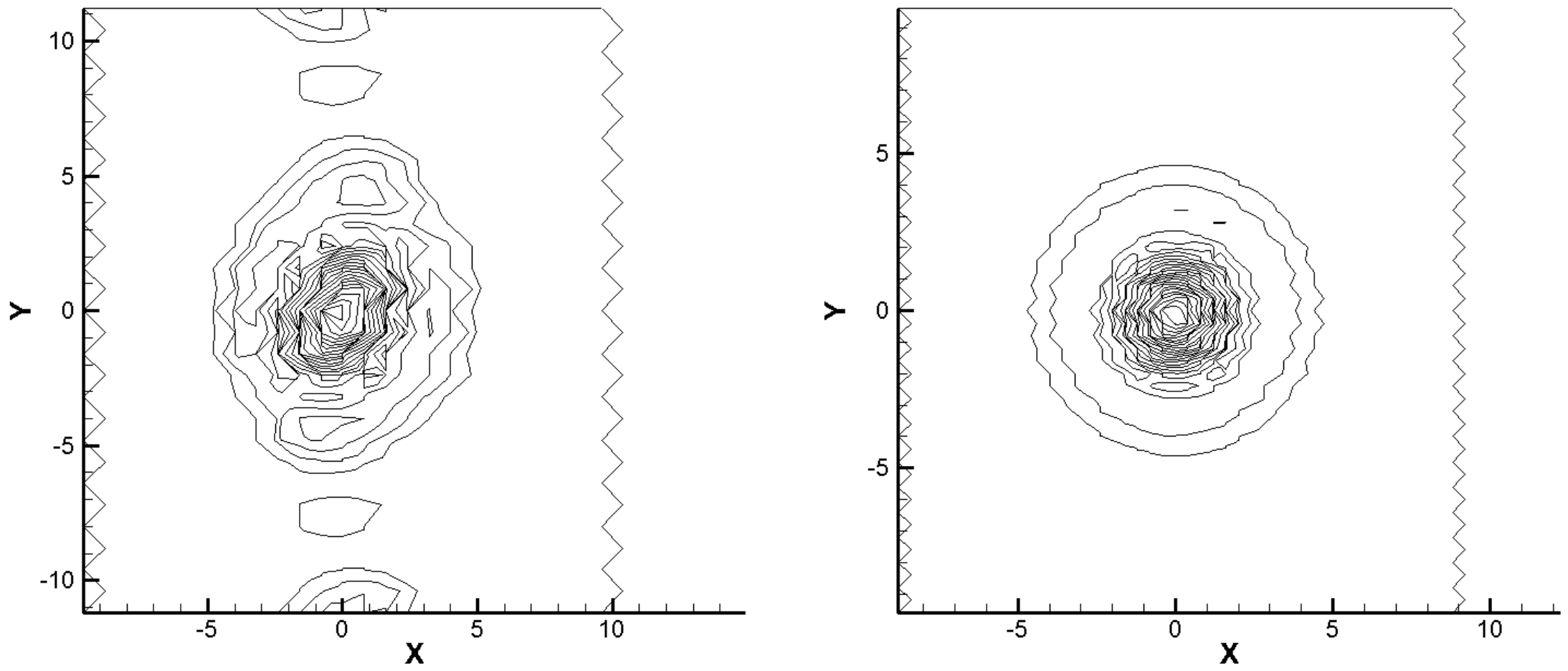

(a) CS4+CF6

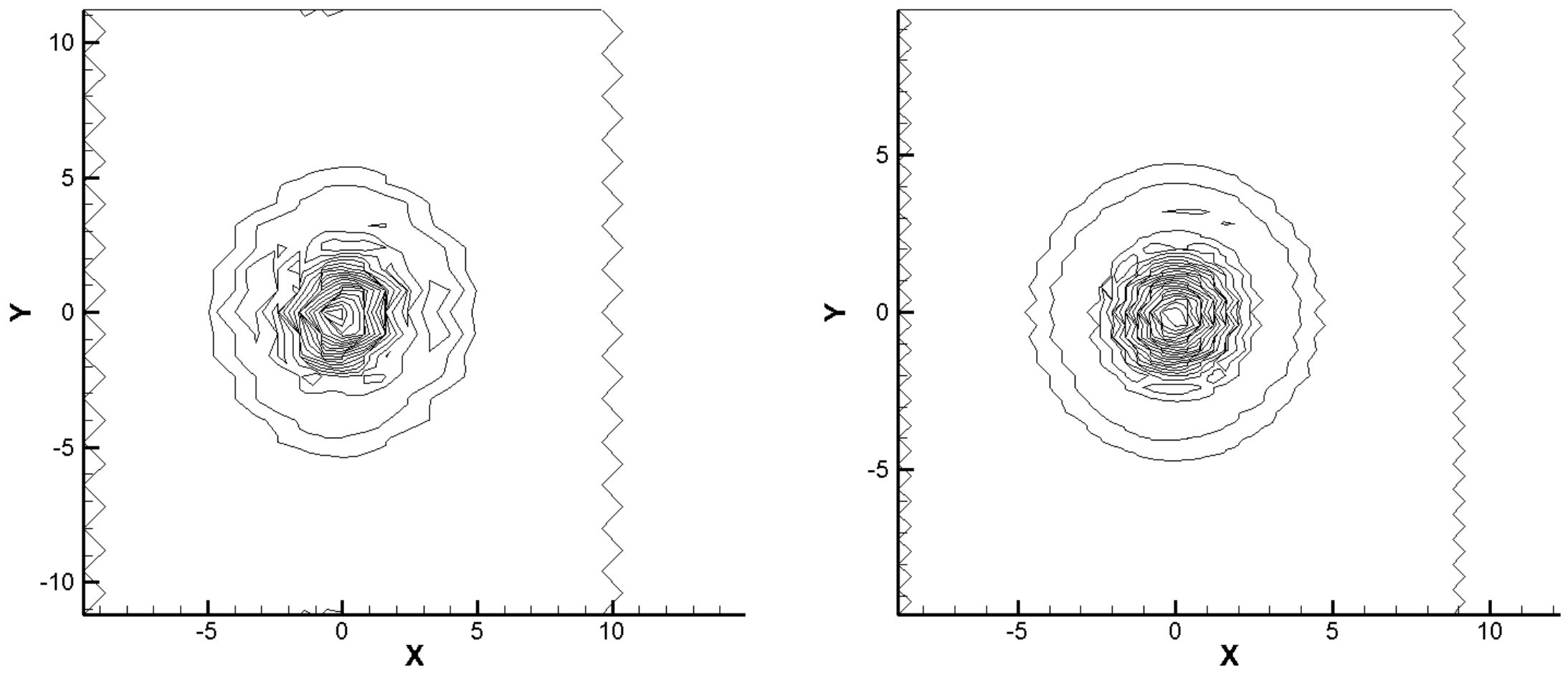

(b) M-UPW3

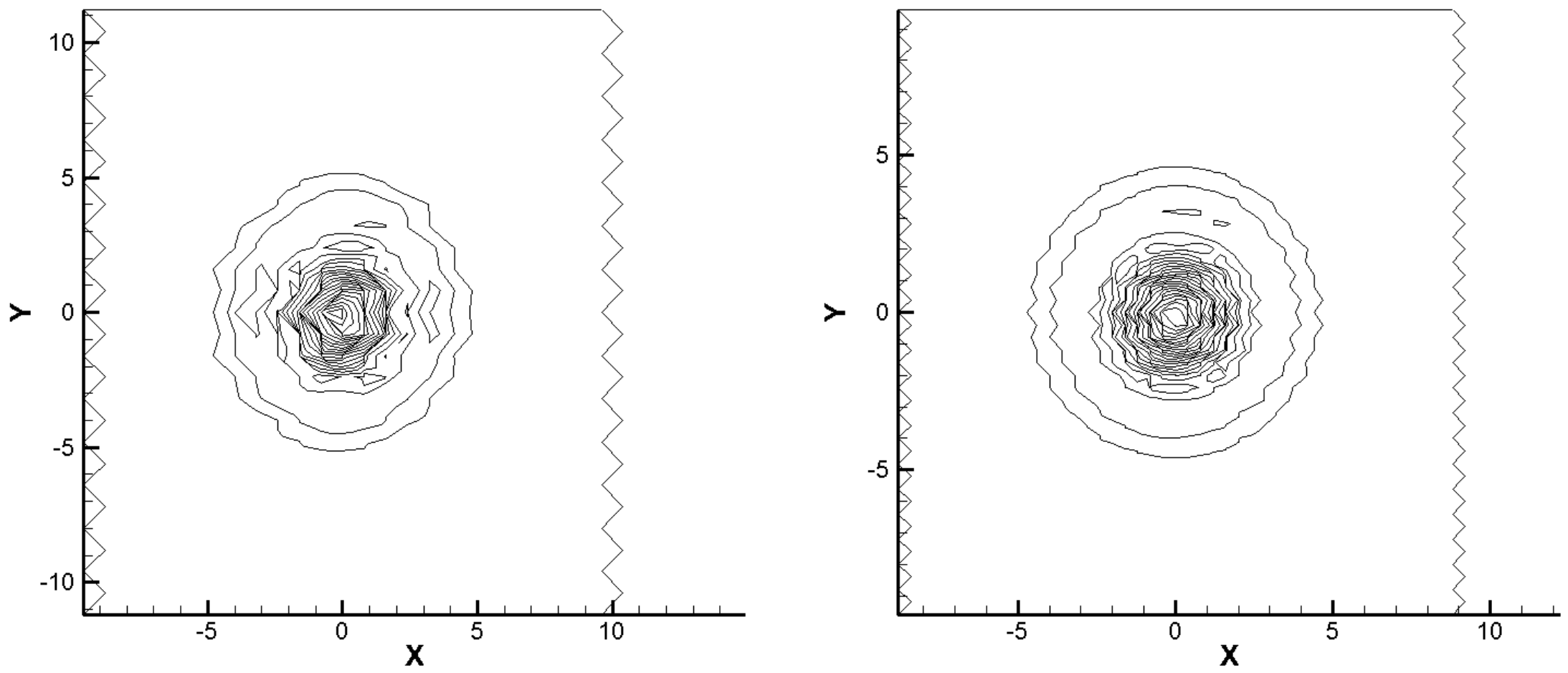

(c) UPW5

Fig. 6. Vorticity contours on triangular grids with the number (41×21) (left) and (81×41) (right) in moving vortex problem

In a quantitative perspective, distributions of the velocity *v*-component on two grids are drawn along the middle horizontal line with $j=J_{max}/2+1$ in Fig.7. On the coarse grid, two upwind schemes show a sharper description than that by CS4+CF6, while the difference become less visible as expected when grid number increases. Considering vorticity contours in Fig.6, it seems that upwind schemes indicate a relative better performance than the central scheme on the coarse grid.

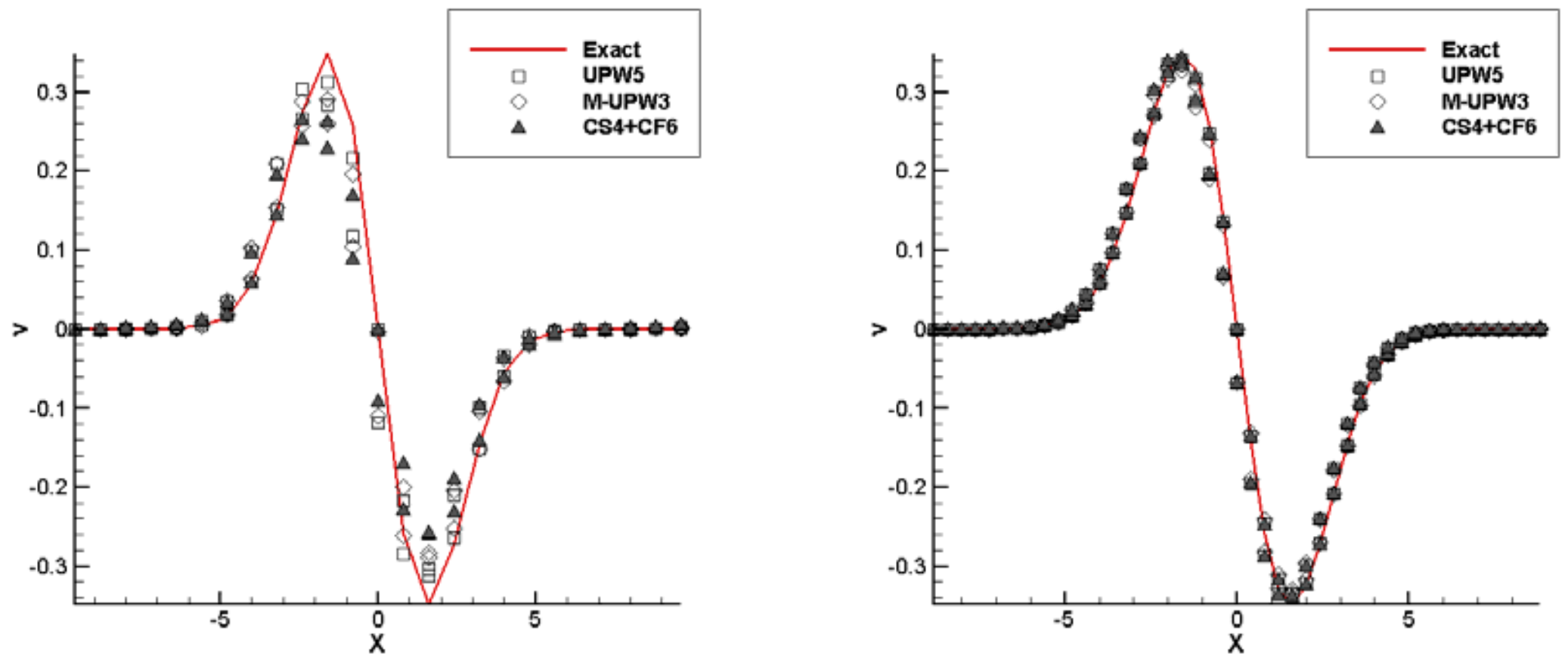

(a) Grid number (41x21) (b) Grid number (81x41)

Fig. 7. Distributions of v-component along the line with $j=J_{max}/2+1$ on triangular grids in moving vortex problem

## 5. Conclusions

In this study, the topic to attain *FSP* for arbitrary upwind schemes is investigated. Although it is known that upwind schemes can acquire *MI*, they are thought to be difficult to achieve *FSP* due to the presence of flux splitting. After careful analysis, the following methods are proposed:

(1) A central scheme decomposition (CSD) method is developed, through which a central scheme called $\delta^{c,(1)}$ is acquired from the upwind scheme for metric evaluations.

(2) Lax-Friedrichs-type splitting scheme is proposed for flux evaluations to combine with the upwind scheme.

(3) Using above methods and the metric forms derived by Thomas, Lombard and Neier [6, 7], *FSP* is achievable for arbitrary upwind node schemes. For half-node or mixed type scheme, an idea of directionally consistent interpolation (DCI) should be employed; otherwise *MI* and *FSP* thereafter will still be violated.

Two cases are chosen for numerical validations, i.e., the problem of *FSP* and that of vortex preservation. Three deformed grids are chosen as wavy grids, seriously randomized grids and triangular grids. Numerical results validate the theoretical outcomes, and the capability of upwind schemes on largely deformed grids is manifested. The computations on triangular grids indicate, if the topology of unstructured grids can be explained as the structured one, the methods discussed in this study are supposed to be applicable to unstructured meshes.

Currently, the investigations are fit to linear upwind schemes, through which low speed compressible problems could be applied to. It is conceivable that the methods cannot be directly used for solving problems with strong discontinuities like shocks. Regarding this aspect, the nonlinear technique has been developed under the framework of schemes with the conservative form and corresponding results will be discussed in other publications.